# Fe-MoS$_2$ nanoenzyme with photothermal enhanced enzyme activity for glucose colorimetric detection

Xiaolu Wang，Guiye Shan*

© Center for Advanced Optoelectronic Functional Materials Research and Key Laboratory of UV Light-Emitting Materials and Technology of Ministry of Education, College of Physics, Northeast Normal University, Changchun 130024, China

*Corresponding author.

E-mail address: shangy229@nenu.edu.cn

**Abstract:** With the development of nanotechnology, it has been discovered that some nanomaterials have the activity of mimicking enzymes. This type of inorganic nanomaterial with characteristics similar to natural enzymes is called nanoenzyme. Compared with natural enzymes, nanoenzymes have advantages such as low deactivation, good stability, low production and storage costs, surface modification, and large-scale preparation. They have a wide range of applications in fields such as human health, environmental safety, and biosensing. MoS$_2$ can replace natural peroxidase to catalyze the decomposition of H$_2$O$_2$ and is considered to have peroxidase like activity, making it a typical nanoenzyme material. In addition, the Fenton reaction between Fe$^{2+}$ and H$_2$O$_2$ can also cause H$_2$O$_2$ to decompose. Both MoS$_2$ and Fe$^{2+}$ can cause the decomposition of H$_2$O$_2$, resulting in the production of hydroxyl radicals (·OH). Hydroxyl radicals can catalyze the oxidation of chromogenic substrates 3,3',5,5'-tetramethylbenzidine (TMB), generating blue single electron oxidation products (oxTMB) and characteristic absorption peaks at 652 nm. Based on this characteristic, we doped Fe$^{2+}$ into MoS$_2$ to obtain Fe-MoS$_2$ composite materials. Firstly, natural glucose oxidase is used to decompose glucose into glucose lactone and hydrogen peroxide. Then, Fe-MoS$_2$ composite material is added, and TMB is added dropwise to observe the blue turning of the mixed solution. Based on the above principle, colorimetric detection of glucose can be achieved. In addition, under infrared irradiation, the activity of peroxidase like enzymes in Fe-MoS$_2$ is enhanced. The enzyme kinetic parameters of Fe-MoS$_2$ with different iron contents under different infrared irradiation were experimentally investigated. Fe-MoS$_2$ composite material is used for colorimetric detection of glucose, with good sensitivity and specificity, low detection limit, and large detection range, making it an ideal composite material for colorimetric detection of glucose.

## 1. Introduction

Natural catalase can decompose hydrogen peroxide, and MoS$_2$, as a transition metal sulfide, can also serve as an ideal catalyst for catalyzing the decomposition of H$_2$O$_2$, enabling the decomposition of hydrogen peroxide. Therefore, it is believed that MoS$_2$ has peroxidase like activity, and divalent iron ions can also react with hydrogen peroxide[1]. Therefore, divalent iron ions are doped into molybdenum disulfide to produce Fe-MoS$_2$ composite materials. Fe-MoS$_2$ decomposes H$_2$O$_2$ to produce hydroxyl radicals, which can catalyze the oxidation of chromogenic substrates 3,3 ', 5,5 ' - tetramethylbenzidine (TMB), generating blue single electron oxidation products (oxTMB) and characteristic absorption peaks at 652 nm[2,3]. Inorganic substances with enzymatic activity, such as divalent iron ions and molybdenum disulfide, are called nanoenzymes. As natural enzyme substitutes, nanoenzymes have excellent catalytic activity and are widely used in the field of biochemical analysis[4].

Colorimetric detection is a common analytical technique in the field of nanoenzyme research, which refers to the qualitative absorption detection and quantitative colorimetric detection of oxTMB concentration by detecting the absorption spectrum of oxTMB[5-7]. This technique is often used in practical applications of nanoenzymes to indicate the strength of their properties[8]. The colorimetric detection technique based on solution absorbance and color change can quantitatively analyze the degree of discoloration of TMB[9]. The colorimetric method, as a visual and low-cost detection method, has received widespread attention[10]. The colorimetric detection platform is mainly based on the colorimetric process induced by nanoenzymes[11,12].

Molybdenum disulfide ($MoS_2$) is a crystalline compound, with each molybdenum disulfide molecule composed of one molybdenum atom and two sulfur atoms[13]. In crystals, molybdenum atoms are located at the center and form covalent bonds with two sulfur atoms[14]. Molybdenum disulfide mainly exists in three types of structures: 2H type (most stable, exhibiting semiconductor properties), 3R and 1T types (exhibiting metallic properties); The layered structure of molybdenum disulfide gives it different physical and chemical properties in two directions[15-17]. In a layered structure, the thin layer of molybdenum disulfide is surrounded by a dense layer of sulfur atoms, while the molybdenum atomic layer between the sulfur atomic layers is relatively sparse[18]. This structure gives molybdenum disulfide a large specific surface area and controllable chemical activity, making it potentially valuable for applications in catalysis, electronic devices, and tribology[19]. Fenton reaction is an advanced oxidation technology that refers to the process of adding transition metal catalysts such as $Fe^{2+}$ or $Fe^{3+}$ to $H_2O_2$ to produce hydroxyl radicals[20]. This reaction is usually carried out in neutral or weakly acidic environments[21,22]. $Fe^{2+}$ can react with $H_2O_2$ to produce hydroxyl radicals (·OH), a process known as the Fenton reaction of $Fe^{2+}$[23]. Doping $Fe^{2+}$ can alter the electronic structure and conductivity of $MoS_2$[24-26]. Within a certain range, as the doping concentration increases, the conductivity of $MoS_2$ also improves[27,28]. The composite material doped with $Fe^{2+}$ and $MoS_2$ has a more efficient ability to decompose hydrogen peroxide[29,30].

## 2. Experimental

### 2.1 Experimental materials

Glucose was supplied by Sinopharm Chemical Reagents (Shanghai, China). Hydrogen peroxide ($H_2O_2$, 30%) was supplied by Beijing Chemical Reagent Company (Beijing, China). Glucose oxidase (GOx) was purchased from Shanghai Yuanye. Polyethylene glycol (PEG-2000), 3,3',5,5'-tetramethylbenzidine (TMB), sodium molybdate ($Na_2MoO_4·2H_2O$), thioacetamide ($C_2H_5NS$), glucose, citric acid ($C_6H_8O_7$), potassium dihydrogen phosphate ($KH_2PO_4$), disodium hydrogen phosphate ($Na_2HPO_4·12H_2O$) and ascorbic acid (AA) were purchased from Sinopharm Chem. Reagent Co.,Ltd (Shanghai, China). All chemicals were of analytical degree and used as received without further purification.

### 2.2 Experimental instruments

The morphology of the composite material was captured and analyzed by a scanning electron microscope (FEG-250), and further single-molecule morphology analysis was obtained by scanning with a transmission electron microscope (TEM, JEOL JEM 2100 F). Exploring the changes in enzyme activity under near-infrared illumination, an 808nm laser (FLMM-0808-761-002W) was used to provide infrared radiation. The ultraviolet-visible (UV-vis) absorption spectra were recorded by a Shimadzu UV2600 spectrometer in Japan. Raman spectra were recorded on an HR800 Raman microscope (Horiba-Jobin Yvon Inc, France). X-ray photoelectron spectroscopy (XPS) characterization was obtained from a Thermo. ESCALAB 280. Their electron spin resonance (ESR) spectra were measured with a Bruker EMXNano electron paramagnetic resonance spectrometer (Billerica, MA).

### 2.3 Synthesis of Fe-MoS$_2$

Add 0.48 g of sodium molybdate and 0.76 g of thiourea to a solvent containing 40 ml of DI and 20 mL of polyethylene glycol, and stir on a magnetic stirrer until evenly mixed. Subsequently, transfer the solution to a 50 ml PTFE inner lining and load it into a high-pressure vessel. Place the high-pressure reactor in a 240 °C oven and heat it at high temperature for 24 hours. After the reaction is completed, the reactor is naturally cooled and removed. The black precipitate in the reactor is alternately washed with DI and alcohol, and the powder sample obtained after three centrifugations is vacuum dried at 60 °C for 12 hours to obtain the black molybdenum disulfide powder sample. The poor dispersibility of Fe-$MoS_2$ makes it prone to aggregation in water, which in turn affects catalytic activity. Modifying surfactants on the surface of Fe-$MoS_2$ and performing ultrasonic treatment may improve the dispersibility of Fe-$MoS_2$. Sodium dodecyl sulfate (SDS), as a common surfactant, can be used as an ideal stabilizer to prevent material aggregation due to its van der Waals forces[31]. Therefore, the Fe-MoS$_2$ nanosheets peeled off after ultrasonic treatment can be combined with SDS to obtain dispersed and stable nanosheets for subsequent experiments.

### 2.4 Photothermal enhancement of peroxidase activity in Fe-MoS$_2$

Place Fe-MoS$_2$ suspensions with different iron contents in a colorimetric dish, irradiate them with infrared light of different powers, record the temperature of the liquid in the dish using an infrared imager, and conduct colorimetric analysis on the degree of color change of TMB caused by Fe-MoS$_2$ at different irradiation times to explore the infrared enhancing enzyme activity of Fe-MoS$_2$.

### 2.5 Glucose detection

The colorimetric detection of glucose relies on TMB solution, H$_2$O$_2$ solution, and glucose oxidase. Firstly, glucose oxidase is added dropwise to the glucose solution. Glucose is decomposed into glucolactone and H$_2$O$_2$ under the action of glucose oxidase. Fe-MoS$_2$ powder is sonicated in water to form a uniform suspension, and then the suspension is added dropwise to the mixed solution after glucose decomposition. Observing the solution changing from colorless and transparent to blue, the discolored solution was measured using a UV visible absorption spectrometer. The glucose content can be determined by the intensity of the absorption peak at 652nm. The process of preparing TMB solution is to weigh 86 mg of TMB, dissolve it in 19.5 ml of DMSO (dimethyl sulfoxide), and stir evenly with a stirrer. Subsequently, take 16 ml of DI and slowly inject it along the cup wall into a beaker containing TMB. After fully injecting, stir for 10 minutes. After no white powder has precipitated, place the prepared TMB solution in a small bottle of medicine and store it away from light.

## 3. Results and discussion

### 3.1 Characterization of Fe-MoS$_2$

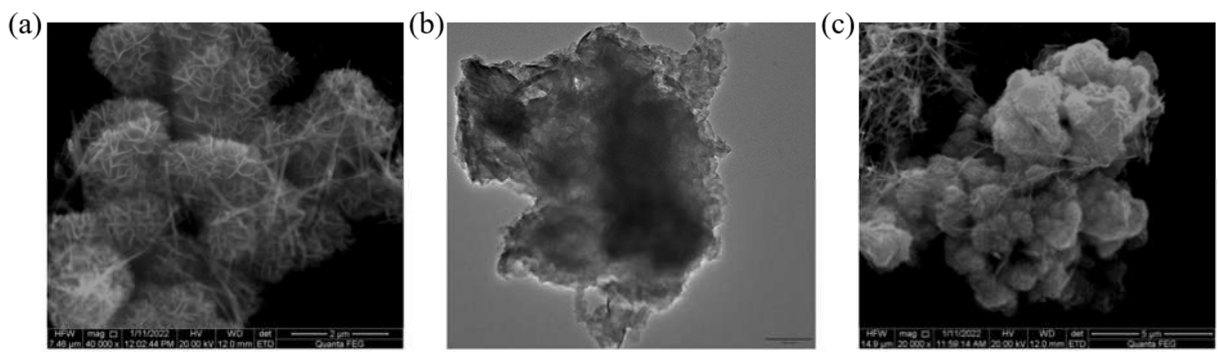

**Figure 1.** (a) SEM images of MoS$_2$ at 2 μm scale; (b) TEM images of MoS$_2$ at 100 nm scale; (c) SEM image of 2.5% Fe-MoS$_2$ at 5 μm scale

From Figure 1(a), it can be seen that MoS$_2$ undergoes different growth rates in various directions during hydrothermal synthesis, resulting in the formation of nanoflower like structures. (b) The TEM image of the figure can clearly observe the microstructure of a piece of MoS$_2$ on the nanopetal obtained by ultrasonic vibration of the nanoflower like product. The diameter of the MoS$_2$ obtained after ultrasonic vibration is about 400 nm. From SEM and TEM images, it can also be visually observed that the nanoflower shaped MoS$_2$ has a large specific surface area and is easy to load onto other carriers.

As shown in Figure 1(c), compared to MoS$_2$, the morphology of Fe-MoS$_2$ remains almost unchanged after iron doping, always exhibiting a nanoflower structure. Moreover, the size of MoS$_2$ slightly increases after the addition of iron, which may be due to the volume expansion of the material caused by the introduction of iron.

Subsequently, Raman spectroscopy was used to characterize the bonding stretching vibration of Fe-MoS$_2$. Raman spectroscopy is a spectral analysis method based on the Raman scattering effect. It utilizes the principle of frequency shift of scattered light after the interaction between laser and matter to obtain molecular vibration information and structural characteristics of matter. The original laser photons undergo Raman scattering by interacting with molecules in the sample.

Scattered light has different frequencies due to its interaction with laser, and the key to Raman spectroscopy is to analyze the sample by measuring the frequency shift and intensity of scattered light.

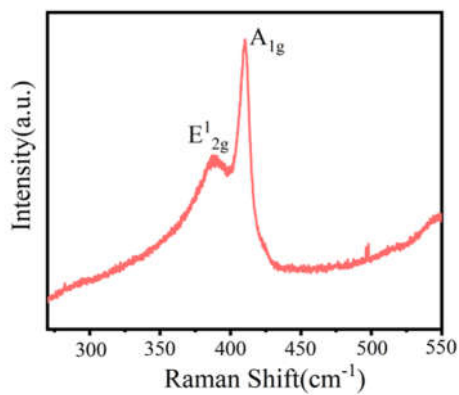

**Figure 2.** Raman spectra of Fe-MoS$_2$, A$_{1g}$ and E$^1_{2g}$ correspond to two Raman active phonon modes of Fe-MoS$_2$, respectively

In Raman spectroscopy, molecular vibration information of the sample can be obtained based on the frequency difference between the laser and the scattered light, such as the stretching, bending, twisting and other vibration modes of chemical bonds. These vibration modes appear in specific peak forms in Raman spectra, with each peak corresponding to a specific vibration mode. By analyzing the intensity and frequency shift of Raman spectra, information on the composition, structure, lattice dynamics, and chemical reaction processes of molecules in the sample can be inferred. Raman mapping (Figure 2) shows MoS$_2$ with typical bands of 2H-MoS$_2$ and 1T-MoS$_2$. The Raman peaks at 380cm$^{-1}$、410 cm$^{-1}$ are the resonance modes of E$^1_{2g}$ and A$_{1g}$ at 488nm laser. E$^1_{2g}$ and A$_{1g}$ are associated with mutually orthogonal atomic shifts, representing in-plane and out-of-plane vibrations respectively, where A$_{1g}$ involves intra-layer displacements of Mo and S atoms. As shown in Figure 2, the Raman characteristic peaks of A$_{1g}$ and E$^1_{2g}$ Fe-MoS$_2$ have appeared, indicating the successful synthesis of Fe-MoS$_2$.

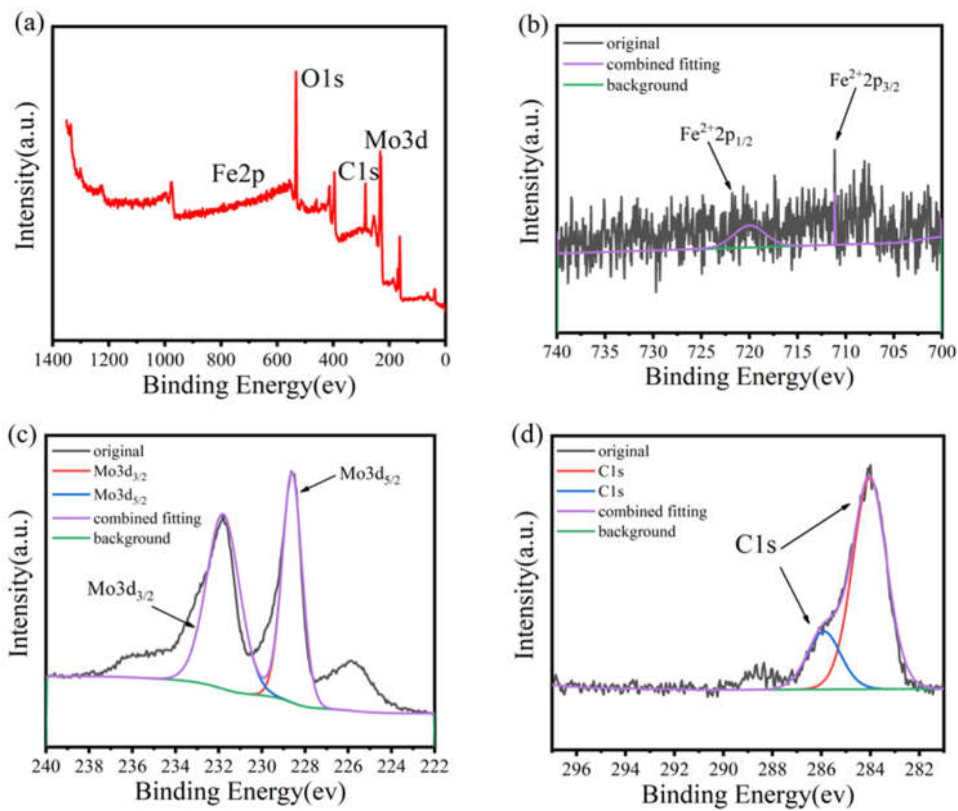

**Figure 3.** XPS analysis spectrum of Fe-MoS$_2$; (a) For the full spectrum; (b) The Fe2p spectrum; (c) Mo3d spectrum; (d) The standard spectrum for C

The element types and valence state distribution of Fe-MoS$_2$ material were analyzed using XPS. In Figure 3 (c), the Mo3d spectrum of the sample showed two peaks near 229 ev and 232 ev, corresponding to the Mo3d$_{5/2}$ and Mo3d$_{3/2}$ components in Fe-MoS$_2$, respectively. Figure 3 (a) demonstrates the presence of Fe, Mo, and S elements in Fe-MoS$_2$, while Figure (b) reflects the valence state of iron ions.

## 3.2 Photothermal enhancement properties of Fe-MoS$_2$

Record the temperature changes of Fe-MoS$_2$ suspension per minute under 808 nm laser irradiation, and create single temperature rise and fall diagrams, multiple temperature rise and fall cycle diagrams, temperature rise diagrams of dispersed solutions with different concentrations, and Fe-MoS$_2$ photothermal heating diagrams with different iron contents. Calculate the photothermal conversion efficiency of Fe-MoS$_2$ with different iron contents.

By using the experiment shown in Figure 4, the temperature rise images of MoS$_2$ and Fe-MoS$_2$ can be obtained under an infrared camera. The infrared camera can record both the infrared photos of the sample in the colorimetric dish and the temperature of the sample liquid. Below, heating and cooling experiments will be conducted on Fe-MoS$_2$ with different iron contents. The temperature of the sample liquid will be recorded during laser irradiation, and it can be seen that the liquid is in a heating process. After turning off the laser, let the liquid cool naturally and record the temperature change of the liquid every minute until it returns to room temperature. Then, turn on the laser light source again and let the laser irradiate the sample liquid for a new round of heating. Repeat this process to obtain the cyclic data shown in Figure 4. The principle for calculating the photothermal conversion efficiency is to irradiate a colorimetric dish containing Fe-MoS$_2$ dispersion with the same power of infrared light. Measure and record physical quantities such as the initial ambient temperature, solution temperature every few minutes, maximum solution temperature, and the time it takes for the solution to cool to the initial temperature after ending illumination, and then substitute them into the formula for calculation.

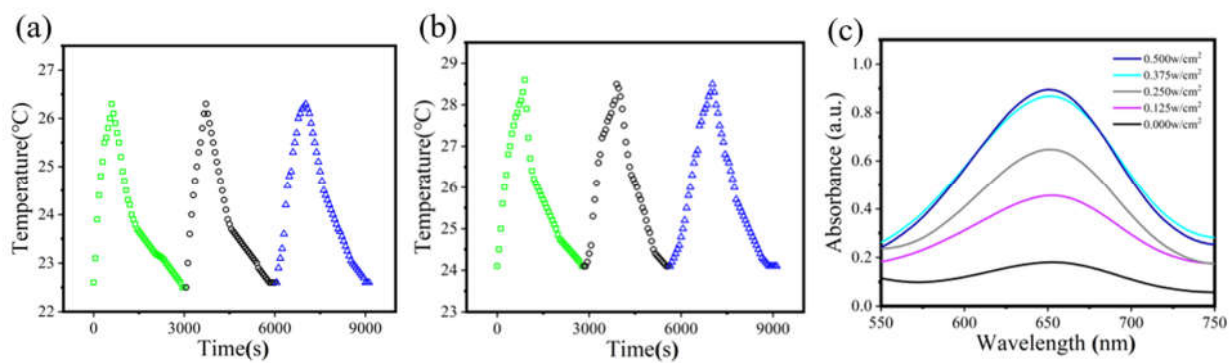

**Figure 4.** (a)(b) Photothermal heating cycle images of 0.5%(a) and 2.5%(b) Fe-MoS$_2$ at 50 μg/ml, respectively.; (c) UV-vis absorption curves of oxTMB with 1.5% Fe-MoS$_2$ and H$_2$O$_2$ added at different laser powers

Figure 4 (a)(b) shows that Fe-MoS$_2$ has good photothermal heating properties. Under infrared light irradiation, the sample itself will heat up. Infrared light enhances the separation effect of electron hole pairs and enhances the activity of peroxidase like enzymes. Each group of experimental samples can reach the same maximum temperature three times and recover to room temperature at the same time. Therefore, Figures 4 (a) and (b) can all indicate that the molybdenum disulfide doped with iron still has good photothermal cycling stability.

**Table 1.** Laser power density measured at 50 μg/ml solution at different laser intensities (0.5cm distance between light source and cuvette)

| I (A) | 0.42 | 0.51 | 0.62 | 0.73 | 0.85 | 1.00 | 1.10 |
|---|---|---|---|---|---|---|---|
| power density (w/cm$^2$) | 0.125 | 0.150 | 0.175 | 0.250 | 0.375 | 0.475 | 0.500 |

Based on the laser power density shown in Table 1, the UV visible absorption spectra of iron doped molybdenum disulfide in the TMB-H$_2$O$_2$ system under different laser power irradiation were measured by changing the laser current, as shown in Figure 4(c).As shown in Figure 4(c), the height of the curve represents the absorption intensity, and the higher the intensity, the stronger the enzyme like activity. From the figure, it can be seen that as the laser irradiation power increases, the absorption peak of oxTMB at 652 nm gradually strengthens, indicating that the enzyme like activity of Fe-MoS$_2$ is stronger. After 0.375w/cm$^2$, increasing the power did not significantly improve the enzymatic activity of Fe-MoS$_2$. This figure intuitively reflects the effect of laser power density on Fe-MoS$_2$ enzyme activity. The results indicate that the catalytic activity of nanoenzymes can be effectively enhanced under a certain laser irradiation power.

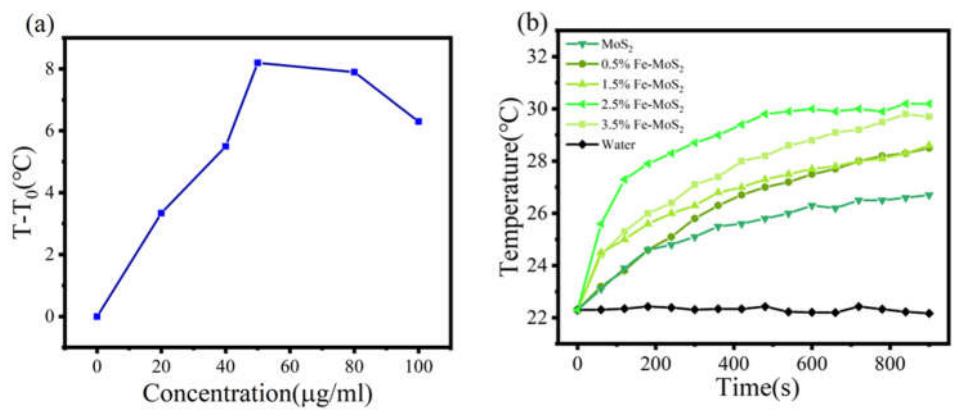

**Figure 5.** (a) The maximum temperature difference diagram of photothermal heating for Fe-MoS$_2$ with different concentrations of 1.5% iron doping; (b) Photothermal heating curves of Fe-MoS$_2$ with 0.5%, 1.5%, 2.5%, and 3.5% iron content (50 µg/ml)

Figure 5 (a) further explores the photothermal heating properties of Fe-MoS$_2$ at different concentrations. The higher the concentration of Fe-MoS$_2$ with photothermal conversion properties, the higher the temperature it can reach under laser irradiation. But when the concentration is high, the dark Fe-MoS$_2$ hinders the propagation of infrared light, and the heating actually weakens. Therefore, based on experimental exploration, the optimal heating concentration of Fe-MoS$_2$ was obtained. Figure 5 (b) explores the photothermal heating properties of Fe-MoS$_2$ with different iron contents.

## 3.3 Fe-MoS$_2$ colorimetric detection of glucose

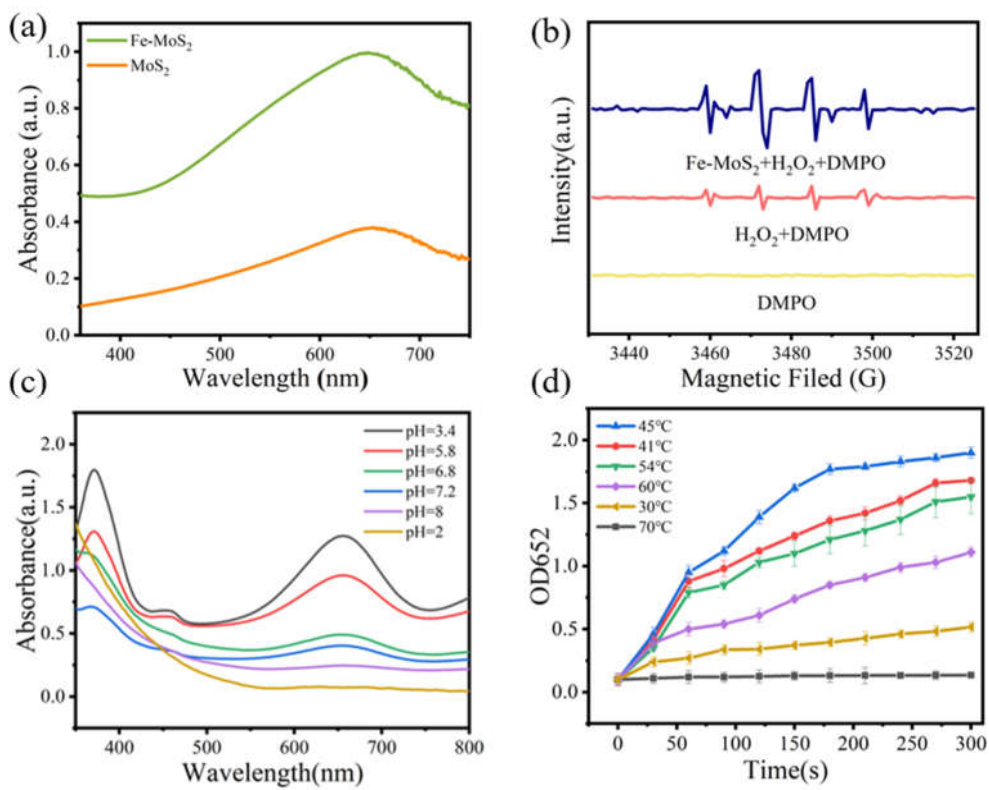

**Figure 6.** (a) UV-vis absorption spectra of MoS$_2$ and Fe-MoS$_2$. UV-vis absorption spectra obtained by placing MoS$_2$ and Fe-MoS$_2$ dispersions of the same concentration in the same TMB-H$_2$O$_2$ system (b) Three curves show the ESR spectra of DMPO in the presence of PBS buffer (blue line), H$_2$O$_2$ (red line), and H$_2$O$_2$+Fe-MoS$_2$ (black line) (c) Fe-MoS$_2$ at different pH values TMB@H$_2$O$_2$ UV-visible absorption spectra of mixed solutions in the system (d) Fe-MoS$_2$@TMB-H$_2$O$_2$ at different temperatures, the OD value of the mixed solution in the system at 652nm

The increase in absorption values and electrochemical signals at characteristic peaks in UV visible absorption spectra is due to the improvement of the interface, and the increase in signal peaks demonstrates the enhancement of effective charge separation and transfer processes. As shown in Figure 6 (a), the characteristic absorption peak of oxTMB is at 652 nm, and the UV visible absorption spectra obtained by placing MoS$_2$ or Fe-MoS$_2$ dispersion in the same TMB-H$_2$O$_2$ system all have characteristic absorption peaks at 652 nm. Moreover, the absorption peak intensity of Fe-MoS$_2$ dispersion is significantly higher at 652 nm, and the absorption curve reflects the strength of the reaction degree. This indicates that Fe-MoS$_2$ has a stronger ability to catalyze the decomposition of H$_2$O$_2$ to produce hydroxyl radicals than MoS$_2$.

As shown in Figure 6 (b), DMPO was used as a free radical scavenger, and the results showed that almost no signal was detected when only DMPO was present. A weak signal peak was obtained in the H$_2$O$_2$+DMPO system, while in the system where H$_2$O$_2$+Fe-MoS$_2$+DMPO existed, four equally spaced signal peaks were captured, with a signal peak intensity ratio of 1:2:2:1. This proves that Fe-MoS$_2$ can catalyze H$_2$O$_2$ to produce hydroxyl radicals, and hydroxyl radicals with strong oxidizing properties are essential for the color oxidation process of TMB.

Figure 6 (c),(d) shows the degree of color change of TMB caused by Fe-MoS$_2$ at different pH values through UV visible absorption spectroscopy testing, and it is found that 3.4 is the pH value with the best nanoenzyme performance. Combining different temperatures Fe-MoS$_2$@H$_2$O$_2$-TMB The system was tested to obtain the OD value of the mixed system at 652nm, and the optimal reaction temperature for the nanoenzyme was determined.

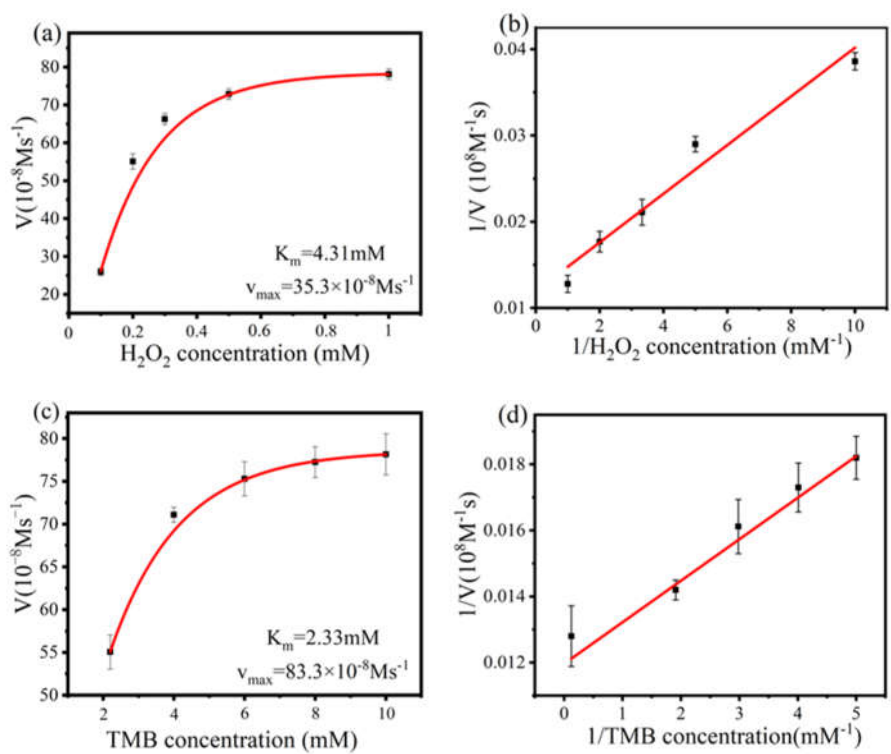

**Figure 7.** (a) Measure the reaction rate at different H$_2$O$_2$ concentrations using Fe-MoS$_2$ (1000 μg/ml) with 0.5% iron content and TMB (10 mM); (b) The fitting plot made using the double reciprocal curve is used to determine the Km value; (c) Measure the reaction rate at different TMB concentrations using Fe-MoS$_2$ (1000 μg/ml) with 0.5% iron content and H$_2$O$_2$ (1 mM); (d) Fit plot made with double reciprocal curve for determining Km value

Through the steady-state kinetic process, a Lineweaver Burk plot can be drawn, with the reciprocal of substrate concentration as the abscissa of the curve and the reciprocal of reaction rate as the ordinate, to draw a fitted straight line. The slope of the fitted line is $K_m/v_{max}$, and the intercept is $1/v_{max}$. From this, the Michaelis Menten constants $K_m$ and $v_{max}$ can be obtained. In the Michaelis Menten equation, the higher the $v_{max}$, the higher the catalytic activity of $H_2O_2$ towards TMB [32,33]; The $K_m$ value is the substrate concentration at which the rate reaches half of the maximum reaction rate, describing the enzyme's affinity for substrates. A lower $K_m$ value indicates a stronger affinity and higher efficiency of the catalyst. Michaelis Menten is the fitting mode for curves, and Lineweaver Burk is the fitting mode for straight lines. After making the experimental data into a scatter plot, curve pattern fitting can be performed on the scatter plot to obtain the corresponding fitting graph of the curve data; Similarly, the fitting of the straight-line scatter plot can be obtained by performing Lineweaver Burk fitting on the straight-line scatter plot obtained by taking the reciprocal of the experimental data.

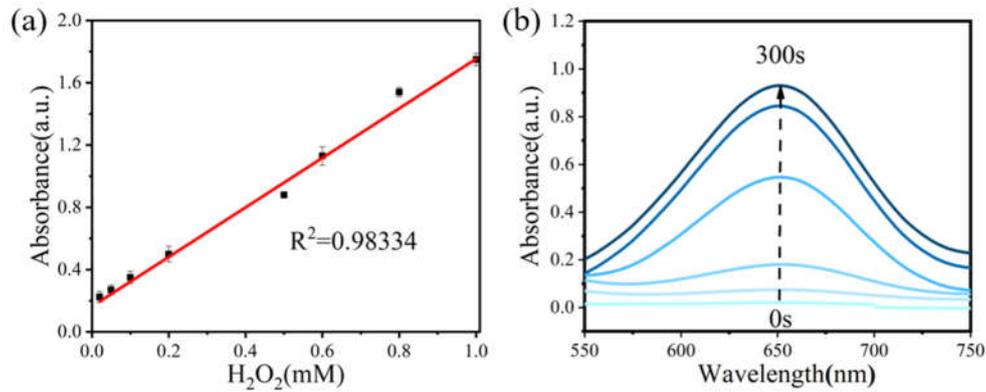

**Figure 8.** (a) UV-vis absorption spectra of TMB discoloration caused by nanoenzymes in different concentrations of $H_2O_2$ at 652 nm absorption value (b) UV-vis absorption spectra of Fe-MoS$_2$ in TMB-H$_2$O$_2$ system for 0-300 seconds

Record the values of the curve and line fitting plots obtained from Figures 7 (a) (b) and 7 (c) (d) to obtain $K_m$, $v_{max}$ values, fitting equations, and variances. These fitting results are summarized in Table 2 to obtain the enzyme kinetic parameters of the composite material. Figure 8 (a) shows the absorption value measured by changing the concentration of hydrogen peroxide and its fitting curve. The detection limit and other parameters of the composite material can be calculated through the straight line in Figure 8 (a)[34]. Figure 8 (b) shows the UV visible absorption curves of the nanoenzyme material and TMB-H$_2$O$_2$ reaction tested every minute from 0 to 300 seconds. It can be seen that the overall value of the curve obtained from each minute of testing is increasing. The reaction rate increases rapidly from the 2nd to the 4th minute, and after 5 minutes, the degree of reaction remains almost unchanged. After measuring the absorption value of UV visible light at 652 nm per minute for the nanoenzyme in the TMB-H$_2$O$_2$ system, the detection range of the nanoenzyme can be calculated. The detection limits of molybdenum disulfide with various iron doping amounts were determined using the absorption values obtained from the reaction of hydrogen peroxide at different concentrations with nanoenzymes. The final results are listed in Table 4.

**Table 2.** The parameters of enzyme kinetics curve for composite materials

| Substrate | $K_m$ /mM | $v_{max}$ /$10^{-8}$ M s$^{-1}$ | Double reciprocal curve | R-Square |
|---|---|---|---|---|
| TMB | 0.0594 | 50.251 | Y=0.01183·X+0.199 | 0.99002 |
| H$_2$O$_2$ | 1.2003 | 47.617 | Y=0.25208·X+0.21001 | 0.97479 |

The above experiment confirms that Fe-MoS$_2$@Cu$_2$O Feasibility of using composite materials for colorimetric detection of glucose. Finally, the photothermal conversion efficiency, detection limit, and detection range corresponding to five different iron contents of molybdenum disulfide are recorded in Table 4. Record the $K_m$ and $v_{max}$ values corresponding to the changes in TMB substrate concentration and H$_2$O$_2$ substrate concentration for five different molybdenum disulfide with different iron contents in Table 3.

Table 3. Kinetic parameters of Fe-MoS$_2$ nanoparticles as catalyst

| Catalyst | TMB | | H$_2$O$_2$ | |
|---|---|---|---|---|
| | Km (mM) | V$_{max}$/10$^{-8}$(Ms$^{-1}$) | Km (mM) | V$_{max}$/10$^{-8}$(Ms$^{-1}$) |
| MoS$_2$ | 0.59 | 21.17 | 1.49 | 19.60 |
| 0.5%Fe- MoS$_2$ | 2.33 | 83.30 | 4.31 | 35.30 |
| 1.5% Fe- MoS$_2$ | 8.40 | 11.20 | 1.21 | 60.20 |
| 2.5% Fe- MoS$_2$ | 2.64 | 15.18 | 7.73 | 43.60 |
| 3.5% Fe- MoS$_2$ | 5.59 | 60.31 | 4.99 | 21.53 |

Through the above series of studies on the properties of peroxidase like enzymes in composite materials, the final data for colorimetric detection of glucose were obtained. The experimental results confirmed the feasibility of colorimetric detection of glucose in composite materials and provided more quantitative references for the application of colorimetric detection of glucose in composite materials.

Table 4. Detection limit, linear detection range, sensitivity, and photothermal conversion efficiency of various catalysts catalyzing hydrogen peroxide

| Nanozymes | LOD(μM) | Linear range(μM) | η |
|---|---|---|---|
| MoS$_2$ | 19.8 | 5.0-500.0 | 13.60% |
| 0.5% Fe-MoS$_2$ | 21.2 | 3.3-540.0 | 14.22% |
| 1.5% Fe-MoS$_2$ | 17.3 | 1.2-590.0 | 15.52% |
| 2.5% Fe-MoS$_2$ | 12.4 | 0.5-600.0 | 16.72% |
| 3.5% Fe-MoS$_2$ | 9.8 | 0.6-590.0 | 14.23% |

## 4.Conclusions

In summary, Fe-MoS$_2$ was prepared by hydrothermal method, and the size of Fe-MoS$_2$ nanoparticles was approximately 400nm, as observed by scanning electron microscopy and transmission electron microscopy. The morphology of Fe-MoS$_2$ is not significantly different from that of MoS$_2$. The successful doping of Fe element can be observed through XPS characterization, which also indicates the presence of elements such as S and Mo. In addition, Fe-MoS$_2$ exhibits enhanced peroxidase like activity under 808nm infrared light irradiation. The detection of glucose plays a crucial role in clinical diagnosis, food analysis and biotechnology. The enhanced catalytic efficiency and substrate selectivity of nanoenzymes will greatly facilitate their practical application. By analyzing the UV visible absorption spectra and calculating the enzyme kinetic parameters of molybdenum disulfide containing iron, it was found that Fe-MoS$_2$ can be applied to colorimetric detection of glucose. This nanoenzyme has high detection sensitivity, low detection limit, and wide detection range, which making it an ideal composite material for colorimetric detection of glucose.